\documentclass{Rinton-P9x6}
  \newcommand{\beq}{\begin{equation}}
  \newcommand{\eeq}{\end{equation}}
  \newcommand{\beql}[1]{\begin{equation}\label{eq:#1}}
  \newcommand{\beqa}{\begin{eqnarray}}
  \newcommand{\eeqa}{\end{eqnarray}}
  \newcommand{\beqas}{\begin{eqnarray*}}
  \newcommand{\eeqas}{\end{eqnarray*}}                       
  \newcommand{\bA}{{\mathbf A}}
  \newcommand{\bC}{{\mathbf C}}
  \newcommand{\bM}{{\mathbf M}}
  \newcommand{\bP}{{\mathbf P}}
  \newcommand{\bS}{{\mathbf S}}
  \newcommand{\bT}{{\mathbf T}}
  \newcommand{\cE}{{\mathcal E}}
  \newcommand{\da}{\dagger}
  \newcommand{\ep}{\epsilon}
  \newcommand{\et}{\eta}

  \newcommand{\ps}{\psi} 
  \newcommand{\rh}{\rho}
  \newcommand{\si}{\sigma} 
  \newcommand{\De}{\Delta}                                          
  \newcommand{\Eq}[1]{Eq.~(\ref{eq:#1})}
  \newcommand{\Tr}{{\mathrm Tr}}
  \newcommand{\eq}[1]{(\ref{eq:#1})}
  \newcommand{\bra}[1]{\langle#1|}
  \newcommand{\ket}[1]{|#1\rangle}
  
  \newcommand{\bracket}[1]{\langle#1\rangle}
  \newcommand{\bx}{{\mathbf x}}
  \newcommand{\cH}{{\mathcal H}}
  \newcommand{\CN}{{\mathrm CN} }

\begin{document}

\title{Quantum Limits of Measurement and Computing Induced by 
Conservation Laws and Uncertainty Relations}

\author{Masanao Ozawa}

\address{Graduate School of Information Sciences,
T\^{o}hoku University, Aoba-ku, Sendai,
980-8579, Japan\\
Mathematical Sciences Research Institute,
1000 Centennial Drive, \#5070, Berkeley, California 94720-5070
}

\maketitle

\abstracts{
A quantitative extension of the Wigner-Araki-Yanase theorem
is obtained on the limitation on precise, non-disturbing 
measurements of observables which do not commute with additive
conserved quantities, and applied to obtaining a limitation on the 
accuracy of quantum computing with computational bases which 
do not commute with angular momentum. }

\section{Introduction}

Errors in quantum computers can be classified into two classes: the
static errors---decoherence in qubits arising from the interaction 
between computational qubits and the environment---and the dynamical 
errors---imperfection of logic operations arising from the interaction
between computational qubits and controllers of quantum gates.   The
current theory of fault-tolerant quantum computing concludes that if
the  imperfection is below a certain threshold, the decoherence can be
corrected in an arbitrarily large quantum computing.\cite{NC00} 
Thus, the  fundamental question as to whether quantum computers are
physically realizable or not can be reduced to such questions as whether
fundamental physical laws lead to an unavoidable imperfection of
quantum logic operations or not.   On the other hand, it has been known
in measurement theory that conservation laws  limit the accuracy of
measurements, as stated by the Wigner-Araki-Yanase (WAY) 
theorem:\cite{Wig52,AY60,91CP,02CUQM}
Observables which do not commute with bounded additive conserved
quantities have no precise, non-disturbing measurements. During
computation, a quantum computer usually needs to ``measure'' internal
qubits to perform the branching program. Such ``measurements''
are carried out by the so-called controlled-not (CNOT) gates.  Thus, it is
natural to ask whether the WAY theorem leads to a conflict
between the accuracy of quantum logic operations and conservation
laws.  

In this paper, we give a quantitative expression of the  WAY theorem
to obtain a lower bound on the error and the disturbance. 
Then, we apply this bound to the implementations of CNOT gates and
obtain the following results: If the computational basis is represented 
by a component of spin and physical implementations obey the angular
momentum conservation law, any physically realizable unitary 
operations on the two computational qubits plus the $S-2$ ancilla
qubits cannot implement the CNOT gate within the error probability 
$1/(4S^{2})$.   An analogous relation for the bosonic  control field is 
also obtained to show that the error probability has the lower bound
$1/(16\bar{n})$, where $\bar{n}$ is the average photon number of the
control field.  From the above result, we can conclude that  in order to
avoid the entanglement between computational qubits and the
controlling system, we need to use the computational basis which 
commute with all conserved quantities, or to use a very large ancilla
or a very strong control field to the extent described by the above 
limitation. The former conclusion is closely related to the proposal of 
``encoded universality''~\cite{BKLW00,DBKBW00}
and the latter is closely related to the recent research on imperfect
operations caused by the entanglement between qubits and the
controlling laser field.\cite{EK01,Gea02}
The present result will give a unified basis for those researches on 
quantum computing, providing with rigorous and general,  
quantitative methods.

\section{Error and Disturbance in Quantum Measurement}

Let $\bM(\bx)$ be a measuring apparatus with macroscopic output
variable $\bx$ to measure, possibly with some error, an observable
$A$  of the {\em object\/} $\bS$, a quantum system represented by
a Hilbert space $\cH_{\bS}$.  The measuring interaction turns on
at time 0 and turns off at time
$\De t$ between the object
$\bS$ and the measuring apparatus $\bM(\bx)$.
We assume that the apparatus has two parts, the {\em probe}
$\bP$ represented by a Hilbert space $\cH_{\bP}$ 
and the {\em ancilla} $\bA$ represented by a Hilbert space
$\cH_{\bA}$, and that the composite system $\bS+\bP+\bA$
evolves unitarily from time 0 to time $\De t$.
Denote by $U$ the unitary operator on
$\cH_{\bS}\otimes\cH_{\bP}\otimes\cH_{\bA}$ representing the time
evolution of $\bS+\bP+\bA$ in the time interval $(0,\Delta t)$. 

At time 0, the object, the probe, and the ancilla are supposed to be in
states  $\psi$, $\phi$, and $\xi$, respectively; all state vectors are
assumed to be normalized unless stated otherwise. 
Thus, the composite system $\bS+\bP+\bA$ is in the state
$\psi\otimes\phi\otimes\xi$ at time 0.  Just after the measuring
interaction turns off, the probe is subjected to a local interaction with the
subsequent stages of the apparatus.  The last process is assumed to
precisely measure an observable $M$ in the probe $\bP$ and to output
the result of the measurement as the value of the macroscopic outcome
variable $\bx$.  
The statistical properties of the apparatus $\bM(\bx)$ is then
determined by the given quadruple $(\cH_{\bP}\otimes\cH_{\bA},
\phi\otimes\xi,U,M)$, which is called the indirect measurement
model of $\bM(\bx)$.\cite{00MN,01OD}
In the Heisenberg picture with the original state 
$\psi\otimes\phi\otimes\xi$, we shall write 
$A(0)=A\otimes I\otimes I$,
$M(0)=I\otimes M\otimes I$,
$A(\Delta t)=U^{\dagger}(A\otimes I\otimes I)U$, 
and $M(\Delta t)=U^{\dagger}(I\otimes M\otimes I)U$. 

We say that measuring apparatus $\bM(\bx)$ {\em measures} 
observable
$A$ {\em precisely}, if   $A(0)$ and $M(\De t)$ have the same
probability distribution on any input  state $\ps$.  
The {\em error operator} $E(A)$ of apparatus 
$\bM(\bx)$ for measuring $A$ is defined by 
$
E(A)=M(\Delta t)-A(0). 
$
The {\em (root-mean-square) error} $\ep(A)$ of 
apparatus $\bM(\bx)$ for
measuring $A$ on input state $\psi$ is, then,  defined by 
$
\ep(A)=\bracket{E(A)^{2}}^{1/2}, 
$ 
where $\langle \cdots\rangle$ stands for 
$\langle \psi\otimes\phi\otimes\xi|\cdots
|\psi\otimes\phi\otimes\xi\rangle$. Then, we have 
\beql{N<E}
\ep(A)^{2}=\si[E(A)]^{2}+\bracket{E(A)}^{2}\ge \si[E(A)]^{2},
\eeq where $\si(X)$ stands for the standard deviation  of an observable
$X$ in $\psi\otimes\phi\otimes\xi$, i.e., 
$ \si(X)^{2}=\langle X^{2}\rangle-\langle X\rangle^{2}.
$
It can be shown that apparatus $\bM(\bx)$ measures observable $A$
precisely if and only if $\ep(A)=0$ for any input state $\psi$.

The apparatus $\bM(\bx)$ is called {\em non-disturbing}, if
$A(0)$ and $A(\De t)$ have the same probability distribution
on any input state $\ps$.
The {\em disturbance operator} $D(A)$ of apparatus 
$\bM(\bx)$ for observable $A$ is defined by 
$
D(A)=A(\Delta t)-A(0).
$
The {\em (root-mean-square) disturbance} $\et(A)$ of 
apparatus $\bM(\bx)$ for observable $A$ on input state $\psi$ is, then, 
defined by 
$
\et(A)=\bracket{D(A)^{2}}^{1/2}. $
Then, we have 
\beql{N<E'}
\et(A)^{2}=\si[D(A)]^{2}+\bracket{D(A)}^{2}\ge\si[D(A)]^{2}.
\eeq
It can be shown that apparatus $\bM(\bx)$ is non-disturbing 
if and only if $\et(A)=0$ for any input state $\psi$.

\section{Quantitative Generalizations of Wigner-Araki-Yanase
Theorem}

Assume the additive conservation law (ACL) on quantities
$L_{1}$ of the object $\bS$, $L_{2}$ of the probe $\bP$,
and $L_{3}$ of the ancilla $\bA$.
In the Heisenberg picture, we shall write
$L_{1}(0)=L_{1}\otimes I\otimes I$,
$L_{1}(\Delta t)=U^{\dagger}(L_{1} \otimes I\otimes I)U$,
$L_{2}(0)=I \otimes L_{2}\otimes I$, 
$L_{2}(\Delta t)=U^{\dagger}(I \otimes L_{2}\otimes I)U$,
and so on.
Then, the ACL is formulated as 
\beql{invariance} 
L_{1}(0)+L_{2}(0)+L_{3}(0) =L_{1}(\Delta t)+L_{2}(\Delta t)+L_{2}(\De t).
\eeq
From \Eq{invariance},
we have the following commutation relations
\beqa\label{eq:noise-commutation1}
[A(0),{L}_{1}(0)]
&=&
[{L}_{1}(\De t),E(A)]+[{L}_{2}(\De t),D(A)]+[{L}_{3}(\De t),D(A)],\\
\label{eq:noise-commutation2}
[A(0),{L}_{1}(0)]
&=&
[{L}_{1}(\De t),E(A)]+[{L}_{2}(\De t),D(A)]+[{L}_{3}(\De t),E(A)].
\eeqa
Taking the modulus of the expectations of the both sides of  
\Eq{noise-commutation1}
and applying the triangular inequality, we have
\beq
|\bracket{[A(0),L_{1}(0)]}|
\le
|\bracket{[{L}_{1}(\De t),E(A)]}|
+|\bracket{[{L}_{2}(\De t),D(A)]}|
+|\bracket{[{L}_{3}(\De t),E(A)]}|.
\eeq
By the Robertson uncertainty relation (i.e., 
$\si(X)\si(Y)\ge|\bracket{[X,Y]}|/2$ for any observables $X$ and $Y$),
we have
\beqa\label{eq:QWAY-1}
\frac{1}{2}|\bracket{\ps|[A,L_{1}]|\ps}|
\le
\ep(A)\si[{L}_{1}(\De t)]
+\et(A)\si[{L}_{2}(\De t)]
+\et(A)\si[{L}_{3}(\De t)].
\eeqa
Similarly, from \Eq{noise-commutation2}, we obtain 
 \beqa\label{eq:QWAY-2}
\frac{1}{2}|\bracket{\ps|[A,L_{1}]|\ps}|
\le
\ep(A)\si[{L}_{1}(\De t)]
+\et(A)\si[{L}_{2}(\De t)]
+\ep(A)\si[{L}_{3}(\De t)].
\eeqa

Each of the above inequalities, \eq{QWAY-1} and \eq{QWAY-2}, 
implies the
WAY theorem as follows.  If the conserved quantities are bounded, we
have
$
\si[{L}_{1}(\De t)],\si[{L}_{2}(\De t)], \si[{L}_{3}(\De t)]<\infty.
$
Thus, if the measurement is precise and no-disturbing, i.e., 
$\ep(A)=\et(A)=0$ for any state $\ps$, then we have
$[A,L_{1}]=0$.
Thus, we conclude that observables which do not commute
with bounded additive conserved quantities allow no precise,
non-disturbing measurement.

Summing up both of the above inequality and using the inequality
$ax+by\le \max\{a,b\}(x+y)$ for $a,b,x,y\ge 0$, we
have
\beq
|\bracket{\ps|[A,L_{1}]|\ps}|
\le
[\ep(A)+\et(A)](2\max\{\si[{L}_{1}(\De t)],\si[{L}_{2}(\De t)]\}
+\si[{L}_{3}(\De t)]).
\eeq
Since $\si(X)\le\|X\|=\|U^{\da}XU\|$ for any observable $X$ and any
unitary operator $U$, by the
inequality 
$(x+y)^{2}\le 2(x^{2}+y^{2})$ we have
\beql{fundamental}
\frac{|\bracket{\ps|[A,L_{1}]|\ps}|^{2}}
{2(2\max\{\|{L}_{1}\|,\|{L}_{2}\|\}+\si[{L}_{3}(\De t)])^{2}}
\le\ep(A)^{2}+\et(A)^{2}.
\eeq

\section{Physical Implementations of CNOT Gates}

Let $U_{\CN}$ be a CNOT gate on a 2-qubit system 
$\bC+\bT$.  
Let $X_{i}$, $Y_{i}$, and $Z_{i}$ be the Pauli operators of qubit $\bC$
for $i=1$ or qubit $\bT$ for $i=2$ defined by
$X_{i}=\ket{1}\bra{0}+\ket{0}\bra{1}$,
$Y_{i}=i\ket{1}\bra{0}-i\ket{0}\bra{1}$, and 
$Z_{i}=\ket{0}\bra{0}-\ket{1}\bra{1}$
with the computational basis
$\{\ket{0},\ket{1}\}$.
On the computational basis, $U_{\CN}$ acts as
$U_{\CN}\ket{a,b}=\ket{a,b\oplus a}$ for $a, b=0,1$,
where $\oplus$ denotes the addition modulo 2.
Thus, in particular, we have
$U_{\CN}\ket{a,0}=\ket{a,a}$
for $a=0,1$.
The above relation shows that the unitary operator $U_{\CN}$
serves as an interaction between the ``object'' $\bC$ and the
``probe'' $\bT$ for a precise, non-disturbing measurement of $Z_{1}$
with probe observable $Z_{2}$ without ancilla system.
Thus, by the WAY theorem, if there
are additive conserved quantities not commuting with $Z_{1}$,
the unitary operator $U_{\CN}$ cannot be implemented correctly.  

Let $\al=(U,\ket{\xi})$ be a physical implementation of 
$U_{\CN}$ defined
by a unitary operator $U$ on the system $\bC+\bT+\bA$, 
where $\bA$ is
the {\em ancilla}, and a state vector
$\ket{\xi}$ of the ancilla.  
The implementation $\al=(U,\ket{\xi})$ defines a 
trace-preserving quantum operation $\cE_{\al}$ by
\beq
\cE_{\al}(\rh)=\Tr_{\bA}[U(\rh\otimes\ket{\xi}\bra{\xi})U^{\da}]
\eeq
for any density operator $\rh$ of the system $\bC+\bT$, 
where $\Tr_{\bA}$ stands for the partial trace over the system $\bA$.
The {\em gate fidelity}~\cite{NC00} of $\al$ is defined by
\beq
F(\cE_{\al},U_{\CN})
=\min_{\ket{\ps}}F(\ps)
\eeq
where $\ket{\ps}$ varies over all state vectors of $\bC+\bT$, and 
$F(\ps)$
is the fidelity of two states 
$U_{\CN}\ket{\ps}$ and $\cE_{\al}(\ket{\ps}\bra{\ps})$.  Then, 
$1-F(\cE_{\al},U_{\CN})^{2}$ is a good measure for the worst
error probability of the implementation $\al$ over all possible
input states.\cite{02CQC}

\section{Imperfection from Angular Momentum Conservation Law}

Let us consider the computational basis defined by the
spin component of the $z$ direction and 
the angular momentum conservation law for the $x$ direction.
Thus, we assume $L_{i}=X_{i}$ for
$i=1,2$, so that  $\|L_{1}\|=\|L_{2}\|=1$,  
and that $L_{3}$ is considered as the $x$-component of 
the total angular momentum divided by $\hbar/2$
of the ancilla system $\bA$.
Then, the unitary operator $U$ should satisfy
the conservation law $[U,L_{1}+L_{2}+L_{3}]=0$. 
Letting $A=Z_{1}$ and $M=Z_{2}$ and applying \Eq{fundamental},  we
have
\beql{squared-noise-bound}
\frac{|\bracket{\ps|[Z_{1},X_{1}]|\ps}|^{2}}
{2[2+\si(L_{3}')]^{2}}
\le\ep(Z_{1})^{2}+\et(Z_{1})^{2},
\eeq 
where $L_{3}'=U^{\da}(I\otimes I\otimes L_{3})U$.

It can be shown that $U=U_{\CN}$ on $\cH_{\bT}\otimes\cH_{\bC}$
if and only if $\ep(Z_{1})^{2}+\et(Z_{1})^{2}=0$ holds for
$\phi=\bracket{0}$ and any $\psi,\xi$.
Thus, $\ep(Z_{1})^{2}+\et(Z_{1})^{2}$ measures the
imperfection of $U$ in implementing $U_{\CN}$. Relation
\eq{squared-noise-bound} suggests that in order to make $U$ more
accurately implement $U_{\CN}$ we need to make $[Z_{1},X_{1}]$ 
smaller or make $\si(L_{3}')$  larger.

In order to relate the measure $\ep(Z_{1})^{2}+\et(Z_{1})^{2}$ to
the gate fidelity and to obtain an upper bound for the gate 
fidelity, assume
$\psi=\frac{1}{\sqrt{2}}(\ket{0}+\ket{1})$
and
$\phi=\ket{0}$.
Then,  we have
\beqa\label{eq:fidelity-bound}
\ep(Z_{1})^{2}+\et(Z_{1})^{2}
&\le&8[1-F(\cE_{\al},U_{\CN})^{2}]
\eeqa
for any $\xi$.\cite{02CQC}
Since $[Z_{1},L_{1}]=[Z_{1},X_{1}]=2iY_{1}$, we have
\beql{commutativity}
|\bracket{[Z_{1},X_{1}]}|=2.
\eeq
Thus, from Eqs.~\eq{squared-noise-bound}--\eq{commutativity}, 
we have the following fundamental upper bound of the gate fidelity
\beq
F(\cE_{\al},U_{\CN})^{2}
\le
1-\frac{1}{4[2+\si(L'_{3})]^{2}}.
\eeq
In the following, we shall interpret the above relation in terms
of the notion of the size of implementations for
fermionic and bosonic ancillae separately.

We now assume that the ancilla $\bA$ comprises qubits.
Then, the size $s(\al)$ of the implementation $\al$ is defined to be
the total number $n$ of the qubits included in $\bC+\bT+\bA$.
Then, we have
$
\De L'_{3}\le \|L_{3}\|=n-2$.
Thus, we have the following upper bound of the gate fidelity
\beql{bound}
F(\cE_{\al},U_{\CN})^{2}\le 1-\frac{1}{4s(\al)^{2}},
\eeq
with $s(\al)=n$.
Therefore, it has been proven that if the computational basis is 
represented by the $z$-component of spin, any implementation
with size $n$ which preserves the $x$-component of angular
momentum cannot implement the CNOT gate within 
the error probability $1/(4n^{2})$.  In particular, any implementation on
$\bC+\bT$ cannot simulate $U_{\CN}$ within the error probability
$1/16$.

In current proposals,\cite{NC00} the external electromagnetic
field prepared  by the laser 
beam is considered to be a feasible candidate for the ancilla 
$\bA$ to be coupled with the computational qubits $\bC+\bT$ via the
dipole interaction.   Then, the ancilla state 
$\xi$ is considered to be a coherent state, for which we have 
$(\De N)^{2}=\bracket{\xi|N|\xi}=\bracket{N}$, where $N$ is the number
operator.  We assume that the beam propagates to the
$x$-direction with right-hand circular polarization.  Then, we
have
$L_{3}=2N$,  and hence
$
\De L'_{3}=2\De N'=2\bracket{N'}^{1/2}\le 2(\bracket{N}+2)^{1/2}$.
Thus, \Eq{bound} holds with defining the size of implementation
$\al$ by
$s(\al)=2\bracket{N}^{1/2}$ appropriately for the strong field, and hence \Eq{bound}
turns to be the relation 
\beql{bound2}
F(\cE_{\al},U_{\CN})^{2}\le 1-\frac{1}{16\bracket{N}}.
\eeq
Therefore, the lower bound of the error probability is inversely proportional 
to the average number of photons.  This bound is consistent with
the recent calculation on error probability based on a ``one-photon type''
transition Hamiltonian.\cite{Gea02}

In this paper, we have concentrated on CNOT gates,  for discussions on
other logic operations we refer to the recent publication.\cite{02CQC}

\section*{Acknowledgments}
This work was supported by the programme ``R\&D on Quantum  
Communication Technology'' of the MPHPT,  by the CREST
project of the JST, and by the Grant-in-Aid for Scientific Research of
the JSPS.

\end{document}